\documentclass[aps,pre,twocolumn,10pt,superscriptaddress,nofootinbib,balancelastpage]{revtex4-1} 
\usepackage[latin1]{inputenc}
\usepackage{amsmath,amssymb}
\usepackage{mathrsfs} 
\usepackage[capitalise]{cleveref}
\usepackage{siunitx}
\usepackage{braket}
\usepackage{calc}
\usepackage{tabularx}
\usepackage{dsfont}
\usepackage{color}
\usepackage{ifthen}
\usepackage[normalem]{ulem}
\usepackage{graphicx}

\allowdisplaybreaks

\newcommand{\ii}{\mathrm{i}}%
\newcommand{\dif}{\mathrm{d}}%
\newcommand{\Nabla}{\vec{\nabla}}%
\newcommand{\Rea}{\operatorname{Re}}%
\newcommand{\uu}{\hat{u}}

\newcommand{\rt}{(\vec{r},t)}
\newcommand{\rvs}{(\vec{r},s)}
\newcommand{\rut}{(\vec{r},\uu,t)}

\newcommand{\ZT}[1]{\textquotedblleft#1\textquotedblright}%

\newcolumntype{Y}{>{\centering\arraybackslash}X}%
\newcolumntype{Z}{>{\raggedright\arraybackslash}X}%

\newlength{\myl}%
\newcommand{\SUM}[2]{{\setlength{\myl}{\widthof{$\displaystyle\sum_{#1}^{#2}$}*\real{0.5}-\widthof{$\displaystyle\sum$}*\real{0.5}}\sum_{#1}^{#2}\;\hspace{-\the\myl}}}
\newcommand{\INT}[3]{\settowidth{\myl}{$\displaystyle\int_{#1}^{#2}$}{\int_{#1}^{#2}\;\;\;\hspace{-\the\myl}\dif #3}\,}
\newcommand{\TINT}[3]{\settowidth{\myl}{$\int_{#1}^{#2}$}{\int_{#1}^{#2}\!\ifthenelse{\equal{#1#2}{}}{}{\;\;\;\;\hspace{-\the\myl}}\dif #3}\,}%
\newcommand{\EINT}[3]{\settowidth{\myl}{$\int_{#1}^{#2}$}{\int_{#1}^{#2}\;\;\;\,\hspace{-\the\myl}\dif #3}\,}

\newcommand{\fn}[2]{#1(#2)}%

\begin{document}
	
\title{Jerky active matter: a phase field crystal model with translational and orientational memory}
\author{Michael te Vrugt}
\affiliation{Institut f\"ur Theoretische Physik, Center for Soft Nanoscience, Westf\"alische Wilhelms-Universit\"at M\"unster, D-48149 M\"unster, Germany}

\author{Julian Jeggle}
\affiliation{Institut f\"ur Theoretische Physik, Center for Soft Nanoscience, Westf\"alische Wilhelms-Universit\"at M\"unster, D-48149 M\"unster, Germany}

\author{Raphael Wittkowski}
\email[Corresponding author: ]{raphael.wittkowski@uni-muenster.de}
\affiliation{Institut f\"ur Theoretische Physik, Center for Soft Nanoscience, Westf\"alische Wilhelms-Universit\"at M\"unster, D-48149 M\"unster, Germany}

\begin{abstract}		
Most field theories for active matter neglect effects of memory and inertia. However, recent experiments have found inertial delay to be important for the motion of self-propelled particles. A major challenge in the theoretical description of these effects, which makes the application of standard methods very difficult, is the fact that orientable particles have both translational and orientational degrees of freedom which do not necessarily relax on the same time scale. In this work, we derive the general mathematical form of a field theory for soft matter systems with two different time scales. This allows to obtain a phase field crystal model for polar (i.e., nonspherical or active) particles with translational and orientational memory. Notably, this theory is of third order in temporal derivatives and can thus be seen as a spatiotemporal jerky dynamics. We obtain the phase diagram of this model, which shows that, unlike in the passive case, the linear stability of the liquid state depends on the damping coefficients. Moreover, we investigate sound waves in active matter. It is found that, in active fluids, there are two different mechanisms for sound propagation. For certain parameter values and sufficiently high frequencies, sound mediated by polarization waves experiences less damping than usual passive sound mediated by pressure waves of the same frequency. By combining the different modes, acoustic frequency filters based on active fluids could be realized.
\end{abstract}
\maketitle

\section{\label{introduction}Introduction}
Almost all field theories for active matter systems are derived in the overdamped limit and neglect the inertia of the active particles. However, there is a current increase of interest in the role of memory and (inertial) delay in active matter \cite{ScholzJLL2018,DauchotD2019,Loewen2020,MijalkovMWV2016,LeymanOWV2018,KhadkaHVC2018,LoosGK2019}. Experiments have shown inertial \cite{ScholzJLL2018,DauchotD2019,Loewen2020} and sensorial \cite{MijalkovMWV2016,LeymanOWV2018} delay to be important for the dynamics of self-propelled particles. Inertia can lead to interesting effects that are not present in overdamped active systems \cite{WagnerHB2019,MandalLL2019,ScholzJLL2018,DauchotD2019,AroldS2020b,Loewen2020,teVrugt2021}, such that inertial active matter models, as presented in Refs.\ \cite{AroldS2020,AroldS2020b}, form an important extension of usual models. A particularly interesting aspect that remains to be explored is the fact that active particles have translational and rotational degrees of freedom which both are associated with inertia, but might relax on different time scales \cite{Loewen2020,Sandoval2020}. Memory and inertia effects are intimately connected and sometimes equivalent, as can be seen from an analysis in the Mori-Zwanzig framework (see \cref{causal}).

The problem of memory effects has been discussed in other contexts, very notably in particle physics \cite{KoideDMK2007,KoideKR2006,Koide2007}. Most dissipative transport equations (such as the diffusion equation) are acausal, since they assume that signals propagate with an infinite velocity. This is typically justified by arguing that microscopic relaxation processes occur infinitely rapidly compared to the relevant dynamics. However, this assumption is not consistent with special relativity where the maximum propagation speed of signals is the speed of light. Relativistic descriptions are required for phase transitions in heavy ion collisions, where the relevant dynamics itself is very fast, such that neglecting memory effects and causality constraints is an approximation that cannot be justified \cite{KoideKR2006}. For this reason, more general theories have been derived that take these effects into account \cite{KoideDMK2007,KoideKR2006,Koide2007,JouCL1999}.

This problem can also arise in soft matter systems such as polymer solutions \cite{JouCL1999}. Although relativistic effects are typically irrelevant here, it is still an approximation to assume that the macroscopic order parameters change much slower than the microscopic degrees of freedom. The time scale will then be set by the damping coefficients rather than by the speed of light. As noted by Archer \cite{Archer2006,Archer2009}, who suggested this as a topic of further investigation within classical dynamical density functional theory (DDFT) \cite{teVrugtLW2020}, the mathematical structure of transport equations for relativistic heavy ion collisions and underdamped soft matter is identical. There is, however, an additional difficulty that can occur in the latter case, namely that two different relaxational time scales are relevant. A typical example would be soft matter systems consisting of particles with orientational degrees of freedom -- in particular active matter -- where position and orientation relax on different time scales. However, different time scales can also be relevant for other systems. Here, we discuss this problem in a very general way and then specialize our results to the case of active field theories.

Phase field crystal (PFC) models \cite{EmmerichEtAl2012}, which provide a coarse-grained description of crystallization and pattern formation in materials, are a particularly suitable framework for such investigations. They were first proposed phenomenologically \cite{ElderKHG2002,ElderG2004,BerryGE2006} and then derived from (dynamical) density functional theory \cite{ElderPBSG2007,vanTeeffelenBVL2009}. The early forms have been extended into a variety of directions including orientational degrees of freedom  \cite{Loewen2010,WittkowskiLB2010,WittkowskiLB2011,WittkowskiLB2011b} and active matter \cite{MenzelL2013,MenzelOL2014}. Moreover, second-order\footnote{By \ZT{$n$-th-order model} we always denote a model that is of $n$-th order in temporal derivatives if not stated otherwise.} models involving inertia have been derived \cite{StefanovicHP2006,StefanovicHP2009,MajaniemiG2007,GalenkoSE2015,HeinonenAKYLA2016}, these are sometimes called \ZT{modified PFC} (MPFC) models \cite{StefanovicHP2006,DehghanM2016}. Consequently, PFC models are a suitable framework for incorporating delay into active field theories, as recently shown by \citet{AroldS2020}. PFC models are reviewed in Ref.\ \cite{EmmerichEtAl2012}, their relation to DDFT is also discussed in Refs.\ \cite{teVrugtLW2020,ArcherRRS2019}.

While these models are of second order in temporal derivatives, memory effects have also been linked to third-order models. A typical example is jerky dynamics, which provides differential equations for the rate of change of the acceleration. These can be relevant in Newtonian mechanics if the force is memory-dependent. The resulting third-order differential equations are useful for obtaining simple models of chaos \cite{EichhornLH1998,EichhornLH2002,Linz1997,Linz1998}. Moreover, third-order dispersions have been studied in optics \cite{TlidiBCHC2013,LeoMKEHT2013}, where they naturally appear in delay systems \cite{SchelteCMGHBSGJG2019}.

In this work, we discuss how to incorporate finite relaxation times in systems where two different time scales are present. This is done in a very general way, since this problem can occur in a variety of contexts. We then specialize to the case of active PFC models and obtain a generalization of the traditional active PFC model \cite{MenzelL2013} that incorporates translational and rotational memory. Notably, the resulting theory is of third order in temporal derivatives and can thus be viewed as a spatiotemporal jerky dynamics. Field theories of this form are almost completely unexplored, and can -- like their aforementioned counterparts from other areas of nonlinear dynamics -- be expected to show very interesting behavior. As shown in the present work, this is relevant for experiments on active matter. In contrast to the passive case, it is found that the damping can affect the linear stability of the fluid state in the PFC model.

An interesting application of the new PFC model is the investigation of sound waves in active matter. In passive systems, sound propagates due to translational inertia. However, it has been found that propagating density waves are also possible in overdamped active systems as a consequence of the coupling between density and polarization \cite{SouslovvZBV2017,Ramaswamy2010}. Our model incorporates both translational inertia and activity. It therefore allows to describe two different mechanisms of sound propagation that are relevant for active matter. In particular, we show that at higher frequencies, the usual \ZT{passive} sound can experience a stronger damping than \ZT{active} sound waves, such that waves of the latter type can propagate further into the medium.

This article is structured as follows: In \cref{causal}, we explain how finite relaxation times can be incorporated into transport equations. We provide a brief introduction to jerky dynamics in \cref{chaos}. A general model for systems with two time scales is developed in \cref{am}. In \cref{pfc}, we use these results to obtain an active PFC model with memory. We perform a linear stability analysis of this model in \cref{linear}. Sound propagation is studied in \cref{sound}. We conclude in \cref{conc}.

\section{\label{causal}Causal field theories}
For a conserved order-parameter field $\phi\rt$, where $\vec{r}$ and $t$ denote position and time, respectively, the time evolution is given by the continuity equation
\begin{equation}
\partial_t\phi\rt=-\Nabla\cdot\vec{J}\rt
\label{continuity}
\end{equation}
with the current $\vec{J}\rt$. A typical example is traditional DDFT \cite{MarconiT1999,MarconiT2000,ArcherE2004,EspanolL2009,teVrugtLW2020}, where the current is
\begin{equation}
\vec{J}\rt=-\tilde{M}\phi\rt\Nabla\frac{\delta F[\phi]}{\delta\phi\rt}
\label{ddft}
\end{equation}
with the mobility $\tilde{M}$ (which for traditional DDFT is given by $\tilde{M} = \beta D$ with the thermodynamic beta $\beta$ and the diffusion constant $D$), the density $\phi$ (which is the order-parameter field), and the free-energy functional $F$. Another example is the Cahn-Hilliard equation \cite{CahnH1958,Cahn1965}, where the current is
\begin{equation}
\vec{J}\rt=-\tilde{M}\Nabla\frac{\delta F[\phi]}{\delta\phi\rt}.
\label{ch}
\end{equation}
A third example is Active Model B \cite{WittkowskiTSAMC2014}, where the current is
\begin{equation}
\vec{J}\rt=-\tilde{M}\Nabla\bigg(\frac{\delta F[\phi]}{\delta\phi\rt} + \lambda(\Nabla\phi\rt)^2\bigg) + \vec{\Lambda}\rt
\end{equation}
with the activity parameter $\lambda$ and the noise $\vec{\Lambda}\rt$. All these currents share a common property: They are \textit{Markovian}, i.e., they only depend on the value of $\phi$ at time $t$, but not on values of $\phi$ at times $s < t$. Moreover, they are \textit{local}, i.e., the time derivative of $\phi$ at position $\vec{r}$ does not depend on the value of $\phi$ at a different position $\vec{r}'$.

This is not the most general case. A good starting point for a systematic analysis is the Mori-Zwanzig formalism \cite{Mori1965,Zwanzig1960,Nakajima1958,Grabert1982,teVrugtW2019,teVrugtW2019d}. It allows to derive macroscopic transport equations for an arbitrary set of relevant variables from the microscopic dynamics and thereby, as discussed in Ref.\ \cite{teVrugtW2019d}, gives very general insights into the general structure of these equations. As exploited in recent work by Meyer \textit{et al.}\ \cite{MeyerVS2017,MeyerPS2020,AmatiMS2019,MeyerWSS2020}, the formalism is also useful for analyzing memory effects in a systematic way. In general, ignoring some degrees of freedom of a system leads to a transport equation in which the dynamics depends on the state of the system at previous times. Approximating this equation by a Markovian equation is possible if the set of relevant variables one has chosen captures the complete macroscopic dynamics \cite{teVrugtW2019d,EspanolL2009}.

Let us assume that we know the microscopic equations governing a passive many-particle system, but are only interested in a conserved scalar order parameter $\phi\rt$ that is a function of the microscopic degrees of freedom. Then, we write the general time evolution as\footnote{In our notation, the operators $\Nabla$ and $\partial_t$ are always understood as acting on the whole term to the right of them. For example, in the expression $\partial_t f(t)g(t) + h(t)$ with time-dependent functions $f$, $g$, and $h$, the operator $\partial_t$ acts on the product $f(t)g(t)$ and not just on $f(t)$. Obviously, $\partial_t$ does not act on $h(t)$.} 
\begin{equation}
\partial_t\phi\rt = \Nabla\cdot\INT{0}{t}{s}\INT{}{}{^3r'}\mathcal{M}(t,s,\vec{r},\vec{r}')\Nabla'\frac{\delta F[\phi]}{\delta\phi(\vec{r}',s)},
\label{conserved}
\end{equation}
where $\mathcal{M}$ is the memory kernel. We have ignored here an organized drift term, which often vanishes for reasons of symmetry if there is only one relevant variable, and we have dropped a noise term. Apart from this, the form \eqref{conserved} is completely general.

The rather complicated general form \eqref{conserved} can be simplified significantly if the relevant variable $\phi$ is slow. In this case, it can be assumed that $\phi$ is constant on the time scales on which the microscopic degrees of freedom relax. For this to be possible, the relaxation of these microscopic degrees of freedom has to occur very rapidly. If we also assume that the nonlocality can be neglected (e.g., because the system is dilute \cite{EspanolL2009}), the exact transport equation \eqref{conserved} can be approximately written as
\begin{equation}
\partial_t\phi\rt = \Nabla\cdot \bigg(\mathcal{D}(\vec{r},t)\Nabla\frac{\delta F[\phi]}{\delta\phi\rt}\bigg)
\label{approx}
\end{equation}
with the diffusion tensor $\mathcal{D}$. This is the so-called \textit{Markovian approximation}.

Although it is made in almost all practical cases, the Markovian approximation is not innocent. From a foundational perspective, it introduces the thermodynamic irreversibility not present in the time-reversal-invariant microscopic laws of physics \cite{teVrugt2020}. In the Markovian limit, an H-theorem corresponding to an increase of entropy can be proven \cite{AneroET2013,WittkowskitVJLB2021}. From a more practical point of view, the Markovian approximation corresponds to the assumption that the set of relevant variables we have chosen gives a complete description of the macroscopic state. For example, if $\phi\rt$ is the number density, making the Markovian approximation implies that the momentum density $\vec{g}\rt$ relaxes very rapidly, i.e., that we are working in the overdamped limit \cite{EspanolL2009}.

What is also relevant here is that \cref{approx} is a diffusive equation which is of first order in temporal and second order in spatial derivatives. This leads, in general, to an infinite propagation speed of signals, i.e., to an \textit{acausal} equation. Early treatments of this problem include inertial extensions of the heat equation \cite{Cattaneo1948} (see Ref.\ \cite{JouCL1999} for a review). This issue is particularly relevant in the case of relativistic systems. If we wish to describe phase transitions in the early universe or in heavy-ion collisions, we generally have to take into account that signals can only propagate with a finite velocity (namely the speed of light) \cite{KoideKR2006}. Thus, when applying the Mori-Zwanzig formalism to relativistic systems, we generally have to be very careful in handling memory effects \cite{KoideDMK2007}. However, as we shall see below, the same problem can arise in soft matter physics, even though the velocities and time scales are very different there.

We start discussing this issue by presenting the theory derived by \citet{KoideKR2006} for describing phase-separation processes in relativistic high energy physics. For a conserved order parameter, the Cahn-Hilliard equation is a very successful theory for phase separation. In order to take into account the finite propagation speed in relativistic systems, the Cahn-Hilliard current \eqref{ch} is modified as
\begin{equation}
\vec{J}\rt = -M\INT{0}{t}{s}e^{-\gamma(t-s)}\Nabla\frac{\delta F[\phi]}{\delta\phi(\vec{r},s)}
\label{mod}
\end{equation}
with the damping coefficient (inverse relaxation time) $\gamma$ and the modified mobility $M=\gamma\tilde{M}$. We have here inserted as the memory kernel a memory function $\gamma\exp(-\gamma(t-s))$, which can be motivated by certain assumptions about the noise. In the Mori-Zwanzig formalism, the memory kernel is related to the correlation of the noise \cite{teVrugtW2019}. The assumption of white noise leads to a Markovian dynamics, whereas colored noise gives the form \eqref{mod} \cite{KoideKR2006}.

The time derivative of \cref{mod} is given by
\begin{equation}
\begin{split}
\partial_t\vec{J}\rt &= - M\Nabla\frac{\delta F[\phi]}{\delta \phi\rt} + \gamma M\INT{0}{t}{s}e^{-\gamma(t-s)}\Nabla\frac{\delta F[\phi]}{\delta\phi(\vec{r},s)}\\
&=- M\Nabla\frac{\delta F[\phi]}{\delta \phi\rt} - \gamma \vec{J}\rt.
\end{split}
\label{rateofchange}\raisetag{2em}
\end{equation}
Differentiating \cref{continuity} with respect to time and inserting \cref{rateofchange} then gives
\begin{equation}
\partial_{t}^2\phi\rt + \gamma \partial_t\phi\rt = M \Nabla^2\frac{\delta F[\phi]}{\delta \phi\rt}.
\label{causalcahnhilliard}
\end{equation}
This is a second-order causal Cahn-Hilliard equation that is similar in form to the telegrapher's equation. Theories of this form can be used to study hyperbolic spinodal decomposition \cite{GalenkoL2008,GalenkoL2008b,KoideKR2007,GalenkoV2007}. 
There are two important limiting cases of \cref{causalcahnhilliard} one can consider. The first one is the overdamped limit $\gamma \to \infty$ with $M/\gamma =\text{const.}$ (corresponding to fixed $\tilde{M}$), in which case we recover the standard Cahn-Hilliard equation
\begin{equation}
\partial_t \phi\rt = \tilde{M}\Nabla^2\frac{\delta F[\phi]}{\delta \phi\rt}.\label{overdampedlimit}
\end{equation}
The second one is the underdamped limit $\gamma \to 0$ at fixed $M$, which gives the inertial Cahn-Hilliard equation
\begin{equation}
\partial_t^2 \phi\rt = M\Nabla^2\frac{\delta F[\phi]}{\delta \phi\rt}.\label{underdampedlimit}
\end{equation}

For the DDFT current \eqref{ddft}, Archer \cite{Archer2006,Archer2009} has noted that this procedure leads to a causal DDFT that has the same structure as the DDFT 
\begin{equation}
\partial_{t}^2\phi\rt + \gamma \partial_t\phi\rt = \frac{1}{m}\Nabla\cdot\bigg(\phi\rt\Nabla\frac{\delta F[\phi]}{\delta \phi\rt}\bigg) 
\label{atomic}
\end{equation}
with particle mass $m$ for particles with inertia. A similar result was obtained by \citet{Chavanis2008}. Within our above considerations about the Mori-Zwanzig formalism, we can give a physical explanation: For systems where inertia is relevant, such as atomic fluids, it is no longer possible to assume that the number density is the only degree of freedom that is relevant, such that we also require the momentum density. Hence, if we want to derive a transport equation for a system with inertia in which the number density is the only relevant variable (i.e., a DDFT for an atomic fluid), we can no longer make the approximation of infinitely fast relaxations. Instead, we need to take into account that in an underdamped system the velocities need a finite time to relax. From \cref{atomic}, we can also see why the over- and underdamped limits have to be taken in the form \eqref{overdampedlimit} and \eqref{underdampedlimit}, respectively: The overdamped limit of \cref{atomic} then leads to traditional DDFT \cite{Archer2009}, whereas the underdamped limit leads to a generalized Euler equation (without convective term). From \cref{atomic}, we find that $M = 1/m$ such that $\tilde{M} = 1/(\gamma m)$. Since we know from DDFT that $\tilde{M}= \beta D$, we can infer $\gamma = 1/(\beta mD)$. Therefore, varying $\gamma$ at fixed $M$ corresponds to changing\footnote{One can also vary $\gamma$ at fixed $M$ by changing $\beta$. However, a change of the temperature will typically also affect the free energy, whereas the diffusion coefficient only appears in the mobility.} $D$, whereas varying $\gamma$ at fixed $\tilde{M}$ corresponds to changing $m$.

Another way to look at this issue is to take into account that equations of motion containing memory effects only arise if we do not consider all degrees of freedom of a system, but only a reduced set \cite{teVrugtW2019d} (Hamilton's equations or the Heisenberg equation of motion have no memory). The memory terms then incorporate (along with the noise) the dynamics of those parts of the system that we do not wish to model explicitly. This also implies that we can, instead of considering the memory, obtain a Markovian dynamics if we enlarge our set of relevant variables \cite{Grabert1982,teVrugtLW2020}. In this case, we can, rather than using a transport equation for the density that contains memory, also write down coupled memoryless equations for mass and momentum density. Mathematically, this is reflected by the fact that the equations with memory are of second order. We could alternatively obtain two equations of motion with the usual first-order structure by using $\partial_t\phi$ as an additional relevant variable. The reason why $\partial_t\phi$ appears is that it is no longer negligible if memory terms are relevant. In the example presented here, we would have $\partial_t\phi = -\Nabla\cdot\vec{g}$, such that using $\partial_t\phi$ as a relevant variable is equivalent to adding the momentum density $\vec{g}$ to the set of relevant variables.

\section{\label{chaos}Jerky dynamics}
For analyzing the problem at hand, the theory of jerky dynamics \cite{Linz1998,Linz1997,EichhornLH2002,Linz2000,Gottlieb2004,Hu2008} will prove to be very useful. A jerky dynamics is an ordinary differential equation of the form \cite{Linz1998}
\begin{equation}
\dddot{x}=J(x,\dot{x},\ddot{x})
\label{jerk}
\end{equation}
with a time-dependent variable $x(t)$ and a function $J$. 
Jerky dynamics is very important for chaos theory \cite{EichhornLH2002,Linz2000}. Mechanically, the \ZT{jerk} is the rate of change of acceleration \cite{Schot1978}. Intuitively, one might expect jerks to be of no importance in classical mechanics, since Newton's equation of motion
\begin{equation}
\ddot{x} = \frac{1}{m}F(x,\dot{x})
\label{newton}
\end{equation}
with the force $F$ is of second order. (Although a jerky dynamics can obviously be obtained by taking the time derivative of \cref{newton}, it would not provide any physical insights.) However, if the force has memory, i.e., if it depends on values of $x$ or $\dot{x}$ at previous times, taking the time derivative of \cref{newton} can lead to an interesting jerky dynamics that contains additive terms depending solely on $x$. Such models can allow for chaos \cite{Linz1998}. 

This relation to memory is what makes jerky dynamics relevant for the present investigation. What will also be useful is that, in some cases, a three-dimensional dynamical system can be written in the form \eqref{jerk}. An example is Sprott's model R \cite{Sprott1994}
\begin{align}
\dot{x}&=a-y,\label{sprott1}\\
\dot{y}&=b+z,\label{sprott2}\\
\dot{z}&=xy-z\label{sprott3}
\end{align}
with the dynamical variables $x$, $y$, and $z$ and the constants $a$ and $b$, which is a simple model for chaos. The problem is addressed in detail in Ref.\ \cite{EichhornLH1998}, where the conditions under which such a transformation is possible are discussed. A useful strategy \cite{EichhornLH1998} is to calculate the first time derivative of \cref{sprott2,sprott3} and the first and second time derivatives of \cref{sprott1}. This gives seven coupled equations for $x,\dot{x},\ddot{x},\dddot{x},y,\dot{y},\ddot{y},z,\dot{z}$, and $\ddot{z}$. These can be used to eliminate $y,\dot{y},\ddot{y},z,\dot{z}$, and $\ddot{z}$, which gives a closed equation for $\dddot{x}$ that solely depends on $x$, $\dot{x}$, and $\ddot{x}$. For \cref{sprott1,sprott2,sprott3}, one obtains \cite{Linz1997}
\begin{equation}
\dddot{x}= - \ddot{x} -x(a-\dot{x}) -b.    
\end{equation}

\section{\label{am}Soft matter with two time scales}
In systems of active (or nonspherical) particles, one needs to take into account both translational and orientational degrees of freedom. As a toy model for describing the physics of such systems, we use the dynamical equation
\begin{equation}
\partial_t\phi\rt = T\rt+R\rt.
\label{toymodel}
\end{equation}
Later, we interpret $T$ and $R$ as the contributions from translational and orientational degrees of freedom, respectively. However, the considerations in this section apply to any physical system whose dynamics can be written in the form \eqref{toymodel}. We are assuming nothing about the form of $T$ and $R$ here -- it can be conserved or nonconserved, active or passive, and it can depend on $\phi$ as well as on spatial derivatives of $\phi$. The field $\phi$ can be a scalar, a vector, or a tensor. As in \cref{causal}, we generalize \cref{toymodel} towards a causal dynamics with time delay\footnote{It is also possible that, in this step, the form of $T$ and $R$ changes compared to \cref{toymodel}, e.g., due to a rescaling of the mobility for dimensional reasons as discussed in \cref{causal}.}:
\begin{equation}
\partial_t\phi\rt = \INT{0}{t}{s}\big(e^{-\gamma_T(t-s)}T\rvs + e^{-\gamma_R(t-s)}R\rvs\big).
\label{smdelay}
\end{equation}
There is a very important difference to the models known from the literature which we have discussed in \cref{causal}: In general, we cannot assume that translational and orientational degrees of freedom relax on the same time scale. Therefore, we have introduced two different relaxation parameters $\gamma_T$ and $\gamma_R$ for translation and orientation, respectively. This has important consequences for the resulting dynamics. To see this, we calculate the time derivative of \cref{smdelay}, which gives
\begin{equation}
\begin{split}
\partial_{t}^2\phi\rt = & -\INT{0}{t}{s}\big(\gamma_T e^{-\gamma_T(t-s)}T\rvs \\
&+\gamma_R e^{-\gamma_R(t-s)}R\rvs\big)\\
&+T\rt+R\rt.
\label{smdelaytd}
\end{split}
\end{equation}
For $\gamma_T=\gamma_R=\gamma$, we could write \cref{smdelaytd} as
\begin{equation}
\partial_{t}^2\phi\rt = - \gamma \partial_t\phi\rt +T\rt+R\rt.
\end{equation}
In general, however, the sum of the first two terms in \cref{smdelaytd} is not proportional to $\partial_t\phi\rt$, since the translational and rotational contributions appear with different prefactors due to the different time scales on which the contributions change. Therefore, \cref{smdelaytd} cannot be written as a second-order partial differential equation without time convolution. 

Physically, this is due to the fact that we require three rather than two variables for a complete description of the system -- in addition to mass and momentum density the angular momentum density is needed. Therefore, we require a third-order partial differential equation to describe the dynamics of $\phi\rt$ without time convolution. This can be obtained using the procedure introduced in \cref{chaos}.

First, we make the definitions
\begin{align}
x\rt &= \phi\rt,\label{x}\\
y\rt &= \INT{0}{t}{s}e^{-\gamma_T(t-s)}T\rvs,\label{y}\\
z\rt &=\INT{0}{t}{s}e^{-\gamma_R(t-s)}R\rvs\label{z}.
\end{align}
For the remainder of this section, we drop the dependence on $\vec{r}$ since it is not important for the further calculations. With the definitions \eqref{x}-\eqref{z}, we can obtain from \cref{smdelay} the dynamical system
\begin{align}
\dot{x} &= y + z,\label{one}\\
\dot{y} &= -\gamma_T y + T,\label{two}\\
\dot{z} &= -\gamma_R z + R\label{three}.
\end{align}
The problem has thus been reduced to the derivation of the jerky dynamics corresponding to the dynamical system given by \cref{one,two,three}. For this purpose, we compute the derivatives
\begin{align}
\ddot{x} &= \dot{y} + \dot{z},\label{four}\\
\ddot{y} &= -\gamma_T \dot{y} + \dot{T},\label{five}\\
\ddot{z} &= -\gamma_R \dot{z} + \dot{R},\label{six}\\
\dddot{x}&= \ddot{y} + \ddot{z}\label{seven}.
\end{align}
One can now use \cref{one,two,three,four,five,six} to express the unknown variables $\ddot{y}$ and $\ddot{z}$ in \cref{seven} in terms of $x$, $\dot{x}$ and $\ddot{x}$.

From \cref{two,five}, we get
\begin{equation}
\ddot{y} = \gamma_T^2 y - \gamma_T T + \dot{T}\label{eight}.
\end{equation}
Similarly, \cref{three,six} give
\begin{equation}
\ddot{z} = \gamma_R^2 z - \gamma_R R + \dot{R}\label{nine}.
\end{equation}
From \cref{two,three,four}, we get
\begin{equation}
\ddot{x}=-\gamma_T y + T - \gamma_R z + R\label{ten}.  
\end{equation}
\cref{seven,eight,nine} lead to
\begin{equation}
\dddot{x} = \gamma_T^2 y + \gamma_R^2 z - \gamma_T T - \gamma_R R + \dot{T} + \dot{R}\label{eleven}.
\end{equation}
From \cref{one}, we obtain
\begin{equation}
y = \dot{x} -z\label{twelve}.
\end{equation}
Moreover, \cref{ten} gives
\begin{equation}
y = - \frac{1}{\gamma_T}\ddot{x} + \frac{1}{\gamma_T} T -\frac{\gamma_R}{\gamma_T}z+\frac{1}{\gamma_T}R\label{thirteen}.    
\end{equation}
Equating \cref{twelve,thirteen} allows to find
\begin{equation}
z =-\frac{\gamma_T}{\gamma_R - \gamma_T}\dot{x} - \frac{1}{\gamma_R -\gamma_T}(\ddot{x} - T - R)\label{fourteen}.
\end{equation}
At this point, we have made the assumption $\gamma_R \neq \gamma_T$. From \cref{twelve,fourteen}, we get
\begin{equation}
\begin{split}
y &= \dot{x} + \frac{\gamma_T}{\gamma_R - \gamma_T}\dot{x} + \frac{1}{\gamma_R -\gamma_T}(\ddot{x} - T - R)\\
&=\frac{\gamma_R}{\gamma_R - \gamma_T}\dot{x} + \frac{1}{\gamma_R - \gamma_T}(\ddot{x} - T - R)\label{fifteen}.
\end{split}
\end{equation}
Finally, combining \cref{eleven,fourteen,fifteen} leads to
\begin{equation}
\begin{split}
\dddot{x}
&=\frac{\gamma_T^2\gamma_R}{\gamma_R - \gamma_T}\dot{x}+\frac{\gamma_T^2}{\gamma_R - \gamma_T}(\ddot{x} - T - R) - \gamma_T T +\dot{T}\\
&\quad\:\!- \frac{\gamma_R^2\gamma_T}{\gamma_R - \gamma_T}\dot{x} - \frac{\gamma_R^2}{\gamma_R - \gamma_T}(\ddot{x } - T - R) -\gamma_R R + \dot{R},
\end{split}\raisetag{3.5em}
\end{equation}
which can be simplified to the final result
\begin{equation}
\dddot{x}= - (\gamma_T + \gamma_R)\ddot{x} - \gamma_T\gamma_R\dot{x} + \gamma_R T + \gamma_T R + \dot{T}+\dot{R}.
\label{finalresult}%
\end{equation}
It is easily verified that \cref{finalresult} also holds for $\gamma_T = \gamma_R$.

\section{\label{pfc}Jerky active matter}
Phase field crystal (PFC) models are a useful framework for the description of soft and active matter \cite{MenzelL2013,EmmerichEtAl2012,OphausGT2018,OphausKGT2020,HollAT2020,HollAGKOT2020}. They can be obtained as a limiting case of the more complex and more general case of DDFT \cite{vanTeeffelenBVL2009,ArcherRRS2019,MenzelOL2014}. Reviews are given by Refs.\ \cite{EmmerichEtAl2012} (PFC models) and \cite{teVrugtLW2020} (DDFT), both reviews discuss the derivation of PFC models from DDFT. The order parameters are the rescaled density $\psi\rt$ and the polarization $\vec{P}\rt$, which arise through an orientational expansion \cite{teVrugtW2019b,teVrugtW2019c} of the one-body density $\rho\rut$ depending on position $\vec{r}$ and orientation $\uu$.

The active PFC model reads \cite{MenzelL2013,MenzelOL2014}
\begin{align}
\partial_t\psi\rt &= \tilde{M}\Nabla^2\frac{\delta F}{\delta \psi\rt} - \tilde{v}_0 \Nabla\cdot\vec{P}\rt,\label{pfc1}\\
\partial_t\vec{P}\rt &= (\tilde{M}\Nabla^2-\tilde{D}_r)\frac{\delta F}{\delta \vec{P}\rt}- \tilde{v}_0 \Nabla\psi\rt\label{pfc2}
\end{align}
with the free-energy functional $F$, activity parameter (a rescaled self-propulsion velocity) $\tilde{v}_0$, and rotational diffusion constant $\tilde{D}_r$. For generality and later convenience, we have introduced a constant mobility $\tilde{M} > 0$, which is set to one in most treatments of the active PFC model, but which here will be useful for taking under- and overdamped limits. 

A generalization towards time-delay dynamics reads
\begin{align}
\begin{split}
\partial_t\psi\rt\label{psi}&= \INT{0}{t}{s}e^{-\gamma_T(t-s)}\bigg(M\Nabla^2\frac{\delta F}{\delta \psi(\vec{r},s)}  \\
&\quad\:\!- v_0 \Nabla\cdot\vec{P}(\vec{r},s)\bigg),
\end{split}\\
\begin{split}
\partial_t\vec{P}\rt &= \INT{0}{t}{s}e^{-\gamma_T(t-s)}\bigg(M \Nabla^2\frac{\delta F}{\delta \vec{P}(\vec{r},s)} -v_0 \Nabla\psi(\vec{r},s)\bigg)\label{pmemory}\\ 
&\quad\:\!-\INT{0}{t}{s} e^{-\gamma_R(t-s)} D_r\frac{\delta F}{\delta \vec{P}(\vec{r},s)},
\end{split}\raisetag{3em}%
\end{align}
where we have introduced the rescaled coefficients $M = \gamma_T \tilde{M}$, $v_0 = \gamma_T \tilde{v}_0$, and $D_r = \gamma_R \tilde{D}_r$. This ansatz assumes that the terms arising from convection and translational diffusion relax on a time scale $\gamma_T^{-1}$, whereas the term arising from rotational diffusion relaxes on a time scale $\gamma_R^{-1}$. Note that \cref{psi} still has the form of a continuity equation for $\psi\rt$. Since all further steps are formally exact, the transformation to the jerky form therefore does not affect the fact that $\psi\rt$ is conserved.

While \cref{psi} can be treated as a second-order dynamics in the standard way since it only depends on one time scale (see \cref{causal}), \cref{pmemory} gives rise to a third-order dynamics. To connect to the general structure presented in \cref{am}, we identify
\begin{align}
\vec{x}\rt &= 
\vec{P}\rt,\label{xrt}\\
\vec{T}\rt&=
M\Nabla^2\frac{\delta F}{\delta \vec{P}\rt}-v_0\Nabla\psi\rt,\label{trt}\\
\vec{R} \rt&=-D_r\frac{\delta F}{\delta \vec{P}\rt}.\label{rrt}
\end{align}
We combine \cref{xrt,trt,rrt} with \cref{finalresult} to the \textit{jerky active matter model}
\begin{align}
\begin{split}
\partial_{t}^2 \psi\rt &= - \gamma_T \partial_t\psi\rt + M\Nabla^2\frac{\delta F}{\delta \psi\rt} \\
&\quad\:\! - v_0 \Nabla\cdot\vec{P}\rt\label{psidrei},
\end{split}\\
\begin{split}
\partial_{t}^3\vec{P}\rt&= - (\gamma_T + \gamma_R)\partial_{t}^2\vec{P}\rt - \gamma_T\gamma_R\partial_t\vec{P}\rt \\
&\quad\:\! - (\gamma_T + \partial_t)D_r \frac{\delta F}{\delta \vec{P}\rt}\label{pdrei}\\
&\quad\:\! + (\gamma_R + \partial_t)\bigg(M\Nabla^2\frac{\delta F}{\delta \vec{P}\rt} - v_0\Nabla\psi\rt\bigg).
\end{split}\raisetag{5em}%
\end{align}
As a consistency check, we confirm that the dynamics of $\vec{P}$ is of second order for $\gamma_T = \gamma_R=\gamma$. In this case, by the line of argument presented in \cref{causal}, it should read
\begin{equation}
\begin{split}
\partial_t^2\vec{P}\rt &= - \gamma \partial_t \vec{P}\rt + (M\Nabla^2 - D_r)\frac{\delta F}{\delta \vec{P}\rt}\\
&\quad\:\! - v_0\Nabla\psi\rt. 
\label{psecondorder}
\end{split}
\end{equation}
Equation \eqref{pdrei} can, for $\gamma_T = \gamma_R=\gamma$, be written as
\begin{equation}
\begin{split}
\vec{0} &= (\partial_t  + \gamma)\bigg(-\partial_t^2\vec{P}\rt- \gamma \partial_t \vec{P}\rt \\
 &\quad\:\! + (M\Nabla^2 - D_r)\frac{\delta F}{\delta \vec{P}\rt}- v_0\Nabla\psi\rt\bigg). 
\label{pthirdorder}
\end{split}
\end{equation}
If we impose \cref{psecondorder} as an initial condition at $t=0$, \cref{pthirdorder} ensures that \cref{psecondorder} is satisfied at all times, such that we have a second-order dynamics. (The necessity of imposing an additional initial condition arises because a differential equation of third order in temporal derivatives requires an initial condition for the second temporal derivative.)

We can also consider the over- and underdamped limits. The overdamped limit corresponds to $\gamma_T,\gamma_R \to \infty$ with fixed $\tilde{M}$, $\tilde{v}_0$, and $\tilde{D}_r$. Thereby, the standard active PFC model given by \cref{pfc1,pfc2} is recovered. The underdamped limit, on the other hand, corresponds to $\gamma_T,\gamma_R \to 0$ at fixed $M$, $v_0$, and $D_r$. In this case, after integrating \cref{pdrei} over $t$, we find
\begin{align}
\partial_t^2 \psi\rt &= M \Nabla^2 \frac{\delta F}{\delta \psi\rt} - v_0 \Nabla \cdot \vec{P}\rt,\\ 
\partial_t^2 \vec{P}\rt &= (M\Nabla^2-D_r) \frac{\delta F}{\delta \vec{P}\rt} - v_0 \Nabla \psi \rt.
\end{align}
Other limits are also possible. For example, we can consider the case in which translational degrees of freedom are underdamped, whereas rotational degrees of freedom are overdamped. This corresponds to $\gamma_T \to 0$ and $\gamma_R \to \infty$ with keeping $M$, $v_0$, and $\tilde{D}_r$ fixed. We then obtain
\begin{align}
\partial_t^2 \psi \rt &= M \Nabla^2 \frac{\delta F}{\delta \psi\rt} - v_0  \Nabla \cdot \vec{P}\rt,\label{mixed1}\\
\partial_t^2 \vec{P}\rt &= (M\Nabla^2 - \tilde{D}_r\partial_t)\frac{\delta F}{\delta \vec{P}\rt} - v_0 \Nabla \psi \rt\label{mixed2}.
\end{align}

Note that the dynamics of $\vec{P}$, given by \cref{mixed2}, is still of second order in the case of completely overdamped orientational dynamics as long as the translational dynamics is underdamped. This, however, is plausible since the local polarization changes not only due to rotations of the individual particles, but also due to translational motion. 

It is interesting to compare \cref{mixed2} to the polarization dynamics given by Eq.\ (5) in Ref.\ \cite{AroldS2020}, which reads (in our notation)
\begin{equation}
\partial_t \vec{P}\rt =  (C_1\Nabla^2 - \tilde{D}_r)\vec{P} - \tilde{v}_0 \Nabla \psi \rt  
\label{eq5}
\end{equation}
with a constant $C_1$. This equation, which is of first order in temporal derivatives, also arises in the context of an underdamped PFC model. However, \cref{eq5} is derived by neglecting orientational convection and by introducing a gradient term proportional to $(\Nabla\vec{P})^2$ in the free energy. Consequently, the gradient term in \cref{eq5} has a different physical origin than the gradient terms in \cref{pfc2,mixed2}, despite the fact that it looks very similar.

\section{\label{linear}Linear Stability}
In the following, we drop the dependence on space and time and restrict ourselves to one spatial dimension. Following Ref.\ \cite{OphausGT2018}, we use the Swift-Hohenberg \cite{SwiftH1977} free-energy functional
\begin{equation}
F= \INT{}{}{x}\Big(\frac{1}{2}(\psi(\epsilon +(1+ \partial_{x}^2)^2)\psi) + \frac{1}{4}(\psi + \bar{\psi})^4 + \frac{b}{2}P^2\Big)
\label{pfcfreeenergy}
\end{equation}
with the constant coefficients $\epsilon$ (shifted rescaled temperature \cite{HollAT2020}), $\bar{\psi}$ (mean density), and $b$. This leads to
\begin{align} 
\frac{\delta F}{\delta \psi} &= (\epsilon + (1+ 
\partial_{x}^2)^2)\psi + (\psi +\bar{\psi})^3,\label{variation1}\\
\frac{\delta F}{\delta P} &= b P\label{variation2}.
\end{align}

Next, we consider small deviations from a reference state $(\psi, P) = (0,0)$ in the form
\begin{align}
\psi&=\psi_1\exp(\lambda t - \ii kx)\label{ansatz1},\\
P&=P_1\exp(\lambda t - \ii kx)\label{ansatz2}
\end{align}
with amplitudes $\psi_1$ and $P_1$, growth rate $\lambda$, imaginary unit $\ii$, and wave number $k$.
We insert \cref{variation1,variation2} together with the ansatz given by \cref{ansatz1,ansatz2} into \cref{psidrei,pdrei}, linearize, and obtain
\begin{widetext}
\begin{align}
\lambda^2 \psi_1&= - \gamma_T \lambda\psi_1 - M k^2(\epsilon + 3\bar{\psi}^2 + (1-k^2)^2)\psi_1 +\ii v_0 k P_1,\label{eigenvalue1}\\
\lambda^3 P_1&= - \gamma_T\gamma_R\lambda P_1 - (\gamma_T + \gamma_R)\lambda^2 P_1 - (\gamma_T + \lambda)D_r b P_1 - (\gamma_R + \lambda)M k^2 b P_1+\ii v_0(\gamma_R + \lambda)k\psi_1\label{eigenvalue2}.
\end{align}
When writing \cref{eigenvalue1,eigenvalue2} as an eigenvalue problem in the form
\begin{equation}
\lambda^3
\begin{pmatrix}
\psi_1\\
P_1
\end{pmatrix}
=
\underline{M}
\begin{pmatrix}
\psi_1\\
P_1
\end{pmatrix}
\label{matrixequation}
\end{equation}
with the matrix
\begin{equation}
\underline{M} = 
\begin{pmatrix}
-\gamma_T \lambda^2 - M k^2(\epsilon + 3\bar{\psi}^2 + (1-k^2)^2)\lambda & \ii v_0 \lambda k\\
\ii v_0 (\gamma_R + \lambda) k &  -\gamma_T\gamma_R\lambda - (\gamma_T + \gamma_R)\lambda^2 - (\gamma_T + \lambda)D_r b  - (\gamma_R + \lambda)M k^2 b
\end{pmatrix}, 
\end{equation}
we get the characteristic polynomial determining the dispersion\footnote{To be able to write \cref{eigenvalue1,eigenvalue2} as the eigenvalue problem \eqref{matrixequation}, we have multiplied \cref{eigenvalue1} by $\lambda$. In \cref{dispersion}, we have then divided by $\lambda$ again. Although this trick can only be applied for $\lambda \neq 0$, we can easily show that \cref{dispersion} is also correct for $\lambda = 0$: In this case, we have $\lambda^2 = \lambda^3$ such that \cref{eigenvalue1,eigenvalue2} form already an eigenvalue problem whose characteristic polynomial is easily found to be given by \cref{dispersion}.}:
\begin{equation}
\begin{split}
0&=(-\gamma_T \lambda - M k^2(\epsilon + 3\bar{\psi}^2 + (1-k^2)^2) - \lambda^2)(-\gamma_T\gamma_R\lambda - (\gamma_T + \gamma_R)\lambda^2 - (\gamma_T + \lambda)D_r b  - (\gamma_R + \lambda)M k^2 b - \lambda^3) \\
&\quad\:\! + v_0^2 (\gamma_R + \lambda) k^2
\label{dispersion}.
\end{split}
\end{equation}

\end{widetext}
\begin{figure*}[tbh]
\includegraphics[scale=1]{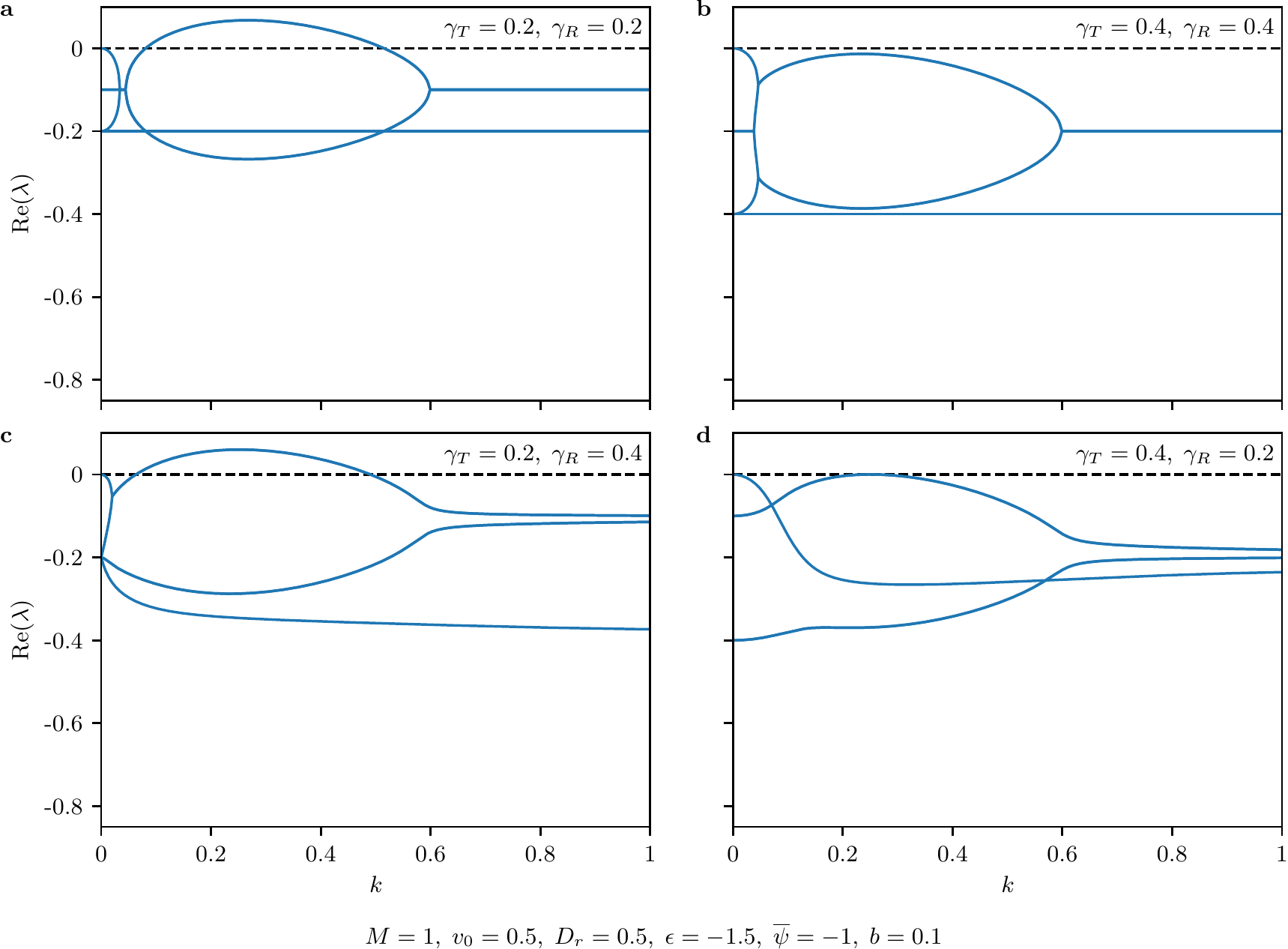}
\caption{\label{fig1}Dispersion relation obtained from a linear stability analysis of the jerky active matter model given by \cref{psidrei,pdrei} for different choices for the values of $\gamma_T$ and $\gamma_R$ and with fixed values for the other parameters. A dashed line indicates $\mathrm{Re}(\lambda)=0$. The homogeneous state is unstable at low and stable at high damping.}
\end{figure*}

To investigate the properties of the jerky active matter model, we started by performing a linear stability analysis of the homogeneous state $(\psi,P)=(0,0)$. This can be done by solving \cref{dispersion} for the complex growth rate $\lambda$ for different values of the real wavenumber $k$. If $\Rea(\lambda) > 0$ for any value of $k$, the system is unstable. Since \cref{dispersion} is a fifth-order polynomial, we have solved it numerically. 

\begin{figure*}[tbh]
\includegraphics[scale=1]{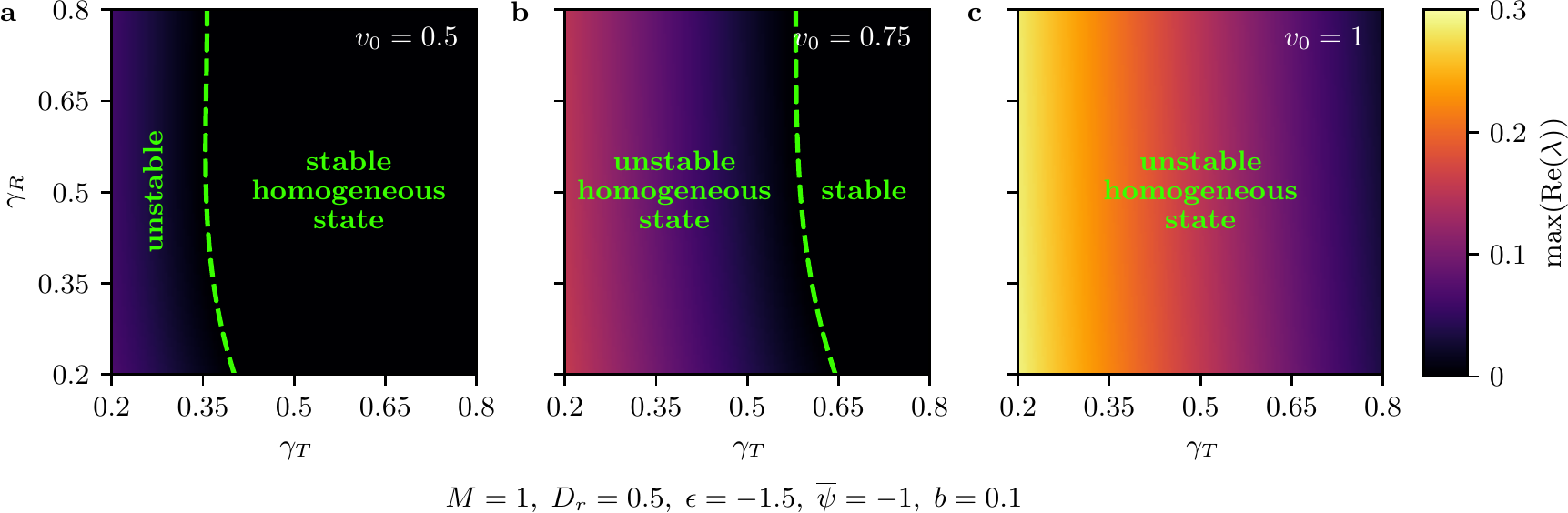}
\caption{\label{fig2}Phase diagram from a linear stability analysis of the jerky active matter model given by \cref{psidrei,pdrei} with variables $\gamma_T$ and $\gamma_R$ for different values of $v_0$ and fixed values of the other parameters. The color indicates the maximum of the real part of $\lambda(k)$. Black regions correspond to a stable, colored regions to an unstable fluid reference state. The phase boundary is indicated as a dashed line. An increased activity $v_0$ destabilizes the system, whereas a larger value of $\gamma_T$ stabilizes it.}
\end{figure*}

The results can be found in \cref{fig1}, which shows the dispersion relation for various values of the damping parameters $\gamma_T$ and $\gamma_R$. The other parameters have been chosen as $M=1$, $v_0 = 0.5$, $D_r = 0.5$, $\epsilon=-1.5$, $\bar{\psi} = -1$, and $b=0.1$, similar to the values used in Ref.\ \cite{OphausGT2018}. There, it was argued that only $\bar{\psi} < 0$ is physical for colloidal systems and that freezing of the liquid state is possible if the temperature parameter $\epsilon$ is negative. In this case, the liquid state is stable for larger $|\bar{\psi}|$.

In our extended model, we find that for $\gamma_T = \gamma_R = 0.2$ (\cref{fig1}a), i.e., for low damping, the homogeneous state is unstable for the given parameters. However, it is stable for $\gamma_T = \gamma_R = 0.4$ (\cref{fig1}b), i.e., for higher damping. This shows that stronger damping can lead to stability of an otherwise unstable state. Consequently, in an active system where inertial delay plays a role, instabilities are more likely to occur at lower damping. One also finds a solution $\lambda = - \gamma$, since for $\gamma_T = \gamma_R = \gamma$, one can (due to \cref{pthirdorder}) factor out a contribution $(\gamma + \lambda)$ in \cref{dispersion}. Since $\gamma > 0$, this does not affect the stability.

A particularly interesting aspect of our model (in which the jerky form is relevant) is the case $\gamma_T \neq \gamma_R$. For $\gamma_T = 0.2$ and $\gamma_R = 0.4$ (\cref{fig1}c), the liquid state is also unstable. Since $\gamma_T = \gamma_R = 0.4$ leads to a stable liquid state, reducing only one damping parameter (in this case $\gamma_T$) while fixing the other one can make the system unstable. Finally, for $\gamma_T = 0.4$ and $\gamma_R = 0.2$ (\cref{fig1}d), the local maximum of Re($\lambda$) also increases slightly above the stability boundary compared to the situation for $\gamma_T = \gamma_R =0.4$ (\cref{fig1}b), but the effect is significantly smaller than in \cref{fig1}c. Consequently, changing $\gamma_R$ has a much smaller effect on the stability of the system than changing $\gamma_T$.

The phase diagram, which is shown in \cref{fig2}, gives a broader picture. It shows the maximum of Re$(\lambda(k))$, i.e., the maximal growth rate, as a function of $\gamma_T$ and $\gamma_R$ for (a) $v_0 = 0.5$, (b) $v_0 = 0.75$, and (c) $v_0 = 1$. 
As can be seen, stronger translational damping (larger $\gamma_T$) stabilizes the homogeneous state, whereas rotational damping (measured by $\gamma_R$) has only weak effects on the linear stability. Moreover, comparing the phase diagrams for different values of $v_0$ shows that activity tends to destabilize the system.\footnote{Note that we fix $M$, $v_0$, and $D_r$ rather than $\tilde{M}$, $\tilde{v}_0$, and $\tilde{D}_r$.}

It is a very remarkable observation that the values of the damping parameters affect the linear stability of the system. In the passive case, \cref{eigenvalue1} leads to the dispersion relation
\begin{equation}
\lambda(k)= -\frac{\gamma_T}{2}\pm \sqrt{\frac{\gamma_T^2}{4}- M k^2 f(k)}   
\label{lambdapassive}
\end{equation}
with the function $f(k)= \epsilon+(1-k^2)^2+3\bar{\psi}^2$ that is independent of $\gamma_T$ and $\gamma_R$. The largest eigenvalue $\lambda$, obtained by choosing the \ZT{+} sign in \cref{lambdapassive}, has a positive real part if and only if $f(k) < 0$ (and $k \neq 0$). 
This reflects the fact that the passive system with damping always approaches the minimum of the free-energy functional $F$, implying that whether the homogeneous state is stable solely depends on whether it is a minimum of $F$. The damping coefficient in the passive case can only determine how and how rapidly equilibrium is approached, but not the equilibrium state itself. The active system, however, does not obey such a minimization principle.\footnote{\citet{AroldS2020}, who also studied underdamped dynamics, have reported that the damping parameter does not affect the state diagram. In our notation, the calculations in Ref.\ \cite{AroldS2020} fix $\tilde{M}$, $\tilde{v}_0$, and $\tilde{D}_r$ rather than $M$, $v_0$, and $D_r$. As discussed in \cref{causal} for the case of DDFT, fixing $M$ corresponds, roughly speaking, to changing the damping coefficient $\gamma = 1/(\beta m D)$ by changing the diffusion coefficient $D$, whereas fixing $\tilde{M}$ corresponds to changing $\gamma$ by changing the mass $m$.}

\section{\label{sound}Sound waves}
\begin{figure*}[tbh]
\includegraphics[scale=1]{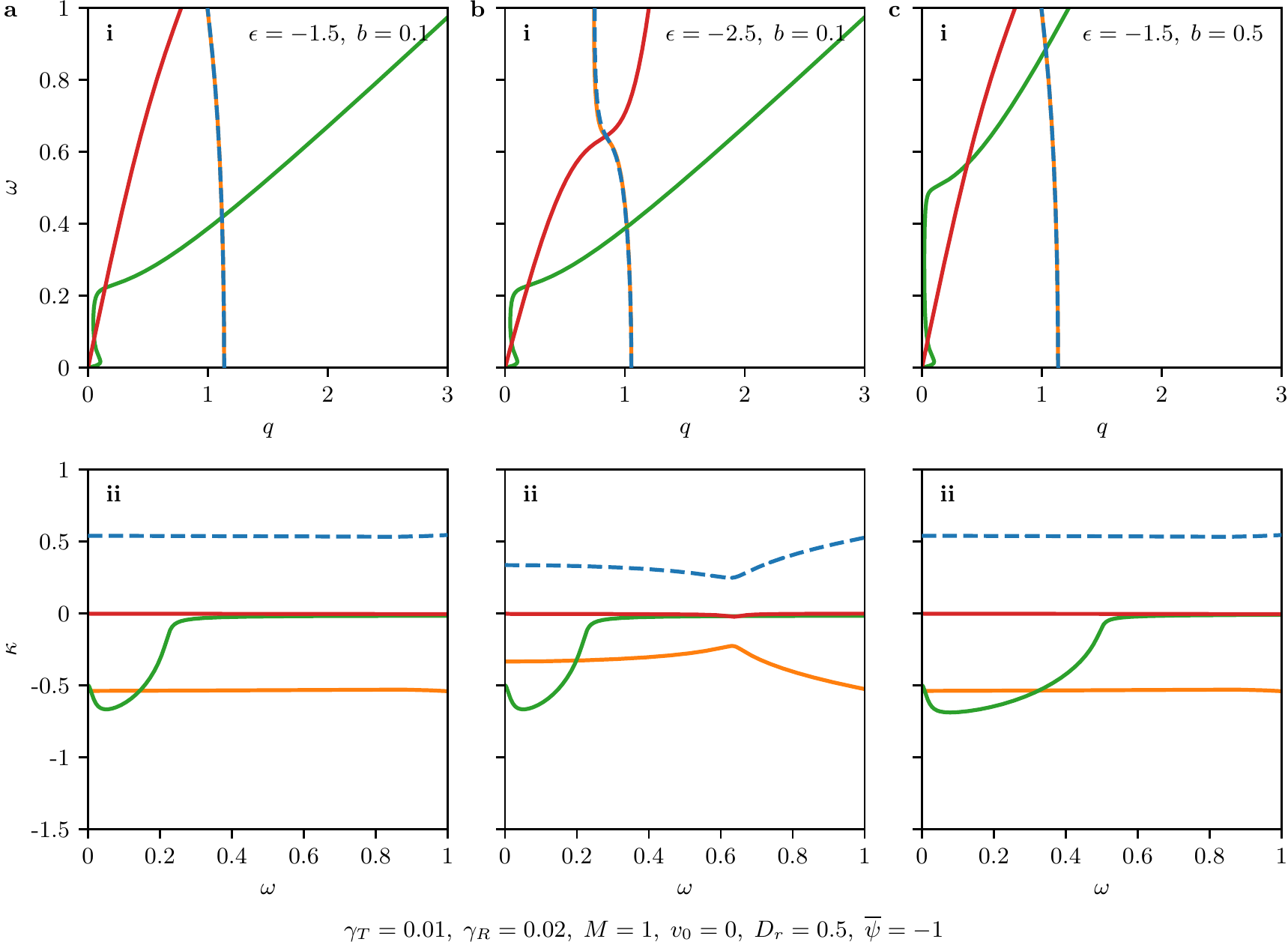}
\caption{\label{fig3}Dispersion relations showing $\omega(q)$ and $\kappa(\omega)$ for sound propagation in a passive system with weak damping. Only modes with positive $q$ and $\omega$ are shown. Different colors are used to distinguish the different modes. The red, blue, and orange modes are unaffected by a change of $b$, the green mode is unaffected by a change of $\epsilon$. For the blue and orange modes, the curves $\omega(q)$ coincide. Since it has positive $\kappa$, the blue mode is shown as a dashed curve.}
\end{figure*}

\begin{figure*}[tbh]
\includegraphics[scale=1]{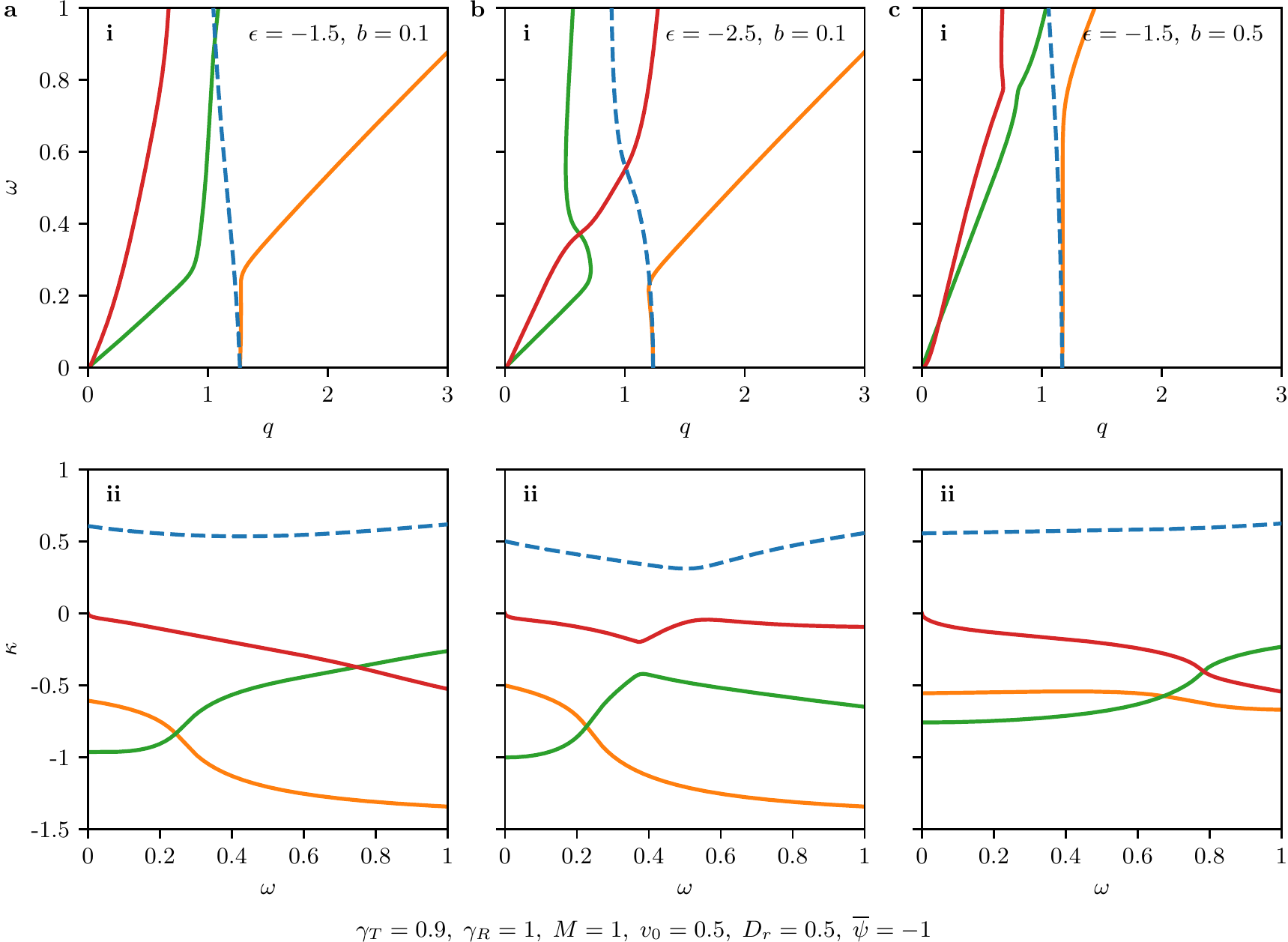}
\caption{\label{fig4}Analogous to Fig.\ \ref{fig3}, but now for sound propagation in an active system with strong damping.
All modes are now affected by changes of $\epsilon$ and $b$.} 
\end{figure*}

Next, we discuss sound propagation, which is a further possible application of underdamped PFC models \cite{StefanovicHP2006,StefanovicHP2009}. Sound waves (propagating density perturbations) in active matter are known to differ from those in passive systems in interesting ways \cite{TuTU1998}. In the passive case, sound waves propagate due to the presence of inertia, and studying them requires hydrodynamic equations in which the momentum density is a relevant variable \cite{Archer2006,WittkowskitVJLB2021}. However, in active systems, sound propagation is also possible in an overdamped system, as can be shown using the Toner-Tu model \cite{TonerT1995}. Here, the density is coupled to a dynamic equation for the polarization, which allows to derive a second-order equation for linear density variations and thus for travelling sound waves \cite{SouslovvZBV2017,Ramaswamy2010}. This effect can also be described in our model, since it also couples the density to the polarization. However, since our model can also incorporate effects of inertia, it still contains sound modes in the passive limit, which then correspond to usual sound waves.\footnote{Strictly speaking, since sound waves are an adiabatic process, one should also include temperature fluctuations \cite{Archer2006}. Interesting insights can, however, already be gained from the isothermal model considered here. For an accurate microscopic description of sound in a nonisothermal system, see Ref.\ \cite{WittkowskitVJLB2021}.} Hence, underdamped active systems, as described by the jerky active matter model, should allow for different types of sound waves.

Sound can also be studied by considering a linear perturbation, i.e., by using the characteristic polynomial \eqref{dispersion}, obtained from the ansatz given by \cref{ansatz1,ansatz2}. However, the physical interpretation and thus the procedure is slightly different: In \cref{linear}, we have solved \cref{dispersion} for the \textit{complex} growth rate $\lambda$, given a fixed \textit{real} value of $k$. Physically, this means that the system is subject to a perturbation $\propto \exp(\ii k x)$ at $t=0$, and we then analyze whether this perturbation grows or decays in time. By considering all values of $k$ (dispersion relation), we then obtain the linear stability, since all linear perturbations can be decomposed in Fourier modes $\exp(\ii k x)$. In the study of sound waves, on the other hand, we are interested in the way in which an oscillation $\exp(\ii \omega t)$ with angular frequency $\omega$ propagates through the system. This implies that we solve \cref{dispersion} for the \textit{complex} wavenumber $k = q + \ii \kappa $ (with $q,\kappa \in \mathbb{R})$, given a fixed $\lambda = \ii \omega$ that is purely \textit{imaginary}. The imaginary part $\kappa$ of $k$ then gives the inverse of the length scale on which the sound wave decays (attenuation length). 

Equation \eqref{dispersion}, which contains only even powers of $k$, then produces four types of solutions, namely (1) $\omega$, $q > 0$, (2) $\omega$, $q < 0$, (3) $\omega > 0$ and $q < 0$, and (4) $\omega < 0$ and $q  > 0$. Here, solutions of types (1) and (2) describe waves travelling to the right (to $+\infty$), whereas solutions of types (3) and (4) describe waves travelling to the left (to $-\infty$). For reasons of symmetry, we can restrict ourselves to solutions of type (1). In this case, solutions with $\kappa < 0$ describe waves that start from $x=0$ and travel to the right with an amplitude that decreases in the direction of propagation. On the other hand, a solution with $\kappa > 0$ would correspond to a wave that travels to the right with an amplitude that increases in the direction of propagation. The latter case is not a reasonable description of a propagating sound wave, such that we focus our discussion on solutions with $\kappa < 0$ (although other solutions are also shown for completeness).

First, we discuss a passive system ($v_0 = 0$) with very weak damping ($\gamma_T = 0.01$, $\gamma_R = 0.02$). This corresponds to the conditions under which sound propagation is usually investigated: For $v_0 = 0$, \cref{psidrei,pdrei} describing $\psi$ and $P$, respectively, decouple. Both equations have travelling waves as a solution. The $\psi$-waves correspond to usual passive sound, i.e., to density perturbations that propagate due to the coupling of density and momentum. On the other hand, the $P$-waves are propagating deviations from the equilibrium polarization. We here assume that the polarization can still be defined\footnote{For an active Brownian sphere, the orientation might only arise through the activity \cite{teVrugtW2019b}, such that there are no orientational degrees of freedom left for $v_0=0$ and $\vec{P}$ cannot be defined. In practice, however, an asymmetry can still be present even if it does not result in self-propulsion. As an example, consider a nonspherical particle that is propelled by ultrasound \cite{VossW2020} -- it still has a (geometrical) orientation if the ultrasound is switched off. If the polarization cannot be defined for $v_0 = 0$, modes linked to the polarization have a physical meaning only for $v_0 \neq 0$.} in a meaningful way for $v_0 = 0$.

The results ($\omega(q)$ and $\kappa(\omega)$) for parameters (a) $\epsilon = -1.5$ and $b= 0.1$, (b) $\epsilon = -2.5$ and $b= 0.1$, and (c) $\epsilon = -1.5$ and $b= 0.5$ are shown in \cref{fig3}, where the other parameters are fixed to $M=1$, $D_r = 0.5$, and $\bar{\psi} = -1$. In the first row, $\omega(q)$ is shown (Figs.\ \ref{fig3}\fn{a}{i}-\ref{fig3}\fn{c}{i}), the second row shows the corresponding curves for $\kappa(\omega)$ (Figs.\ \ref{fig3}\fn{a}{ii}-\ref{fig3}\fn{c}{ii}). Four modes can be found: First, there is a red mode that propagates almost without damping $(\kappa \approx 0)$. Second, there is a green mode for which $q(\omega)$ is small at small $\omega$, followed by a region with stronger growth at larger $\omega$. This mode is damped at small $\omega$ and propagates further into the medium at larger $\omega$. Third, there are a blue mode and an orange mode where $\omega(q)$ has (for these parameters) a strong negative slope and reaches $\omega = 0$ for $q \neq 0$. The damping of the orange mode has no strong dependence on the frequency, whereas the blue mode has positive $\kappa$ and is therefore not a physically reasonable sound wave (we plot it as a dashed line to indicate this).

When comparing the various plots, the physical origin of the different modes becomes transparent: The green mode is identical in \cref{fig3}a and \cref{fig3}b, whereas the other modes show a significant change: In \cref{fig3}\fn{b}{i}, they intersect at $\omega \approx 0.6$, which is not the case in \cref{fig3}\fn{a}{i}. At $\omega \approx 0.6$, one also observes a maximum for the orange and a minimum for the blue mode in \cref{fig3}\fn{b}{ii}, which is not the case in \cref{fig3}\fn{a}{ii}. On the other hand, the red, blue, and orange modes do not change between \cref{fig3}a and \cref{fig3}c, whereas the growth of $q$ with $\omega$ in the green mode now starts at $\omega \approx 0.5$ in \cref{fig3}\fn{c}{i} rather than $\omega \approx 0.2$ as in \cref{fig3}\fn{a}{i}. Since only $\epsilon$, a parameter affecting $\psi$, has been changed between Figs.\ \ref{fig3}a and \ref{fig3}b, it follows that the red, blue, and orange modes correspond to density waves. The red mode is usual passive sound, whereas the orange and blue modes arise from the higher-order spatial gradients in the free energy \eqref{pfcfreeenergy}. (We have verified that the orange and blue modes disappear if these terms are not present in \cref{pfcfreeenergy}.) As far as $\omega(q)$ is concerned, the orange and blue modes coincide, which is not the case for $\kappa(\omega)$. On the other hand, only the parameter $b$, which affects $P$, has been changed between \cref{fig3}a and \cref{fig3}c, such that the green mode corresponds to polarization waves.

The situation is quite different in the active case ($v_0 = 0.5$), which is shown in \cref{fig4}. First, we need to take into account here that, as shown in \cref{linear}, activity tends to destabilize the system. Consequently, we have to increase the damping if we want sound waves in a fluid that does not spontaneously become a crystal. Here, we choose $\gamma_T = 0.9$ and $\gamma_R = 1$. Apart from the change in damping and activity, we use the same parameters as in \cref{fig3}. 

When comparing Figs.\ \ref{fig4}a-\ref{fig4}c, it is found that a change of both $\epsilon$ and $b$ affects all four modes. The reason is that, in the active case, oscillations of $\psi$ and $P$ are coupled. Moreover, the shapes of the modes change. For the red mode, the most significant change in $\omega(q)$ is a sharp bend observed at $\omega \approx 0.8$  in \cref{fig4}\fn{c}{i}. Moreover, the red mode is now damped for larger $\omega$, whereas it can propagate almost without damping at small $\omega$ (\cref{fig4}\fn{a}{ii}-\cref{fig4}\fn{c}{ii}). When considering the green mode, strong variations can be found for both $\omega(q)$ and $\kappa(\omega)$. For $\epsilon = -1.5$ (Figs.\ \ref{fig4}a and \ref{fig4}c), both functions grow monotonously. In contrast, for $\epsilon = -2.5$, $\omega(q)$ is not even a function due to a local maximum in $q(\omega)$  (\cref{fig4}\fn{b}{i}) and $\kappa(\omega)$ reaches a local maximum at $\omega \approx 0.4$ (\cref{fig4}\fn{b}{ii}). A very significant change compared to \cref{fig3} is also observed for the orange mode: Now, $\omega(q)$ has a \ZT{hockey-stick shape}, where for larger frequencies the mode grows to the right (and not to the left as in the passive case). Moreover, $\omega(q)$ does not coincide for the blue and orange mode in the active case.

Consequently, if we define \ZT{sound} as \ZT{propagating density oscillation}, we find two different mechanisms for sound propagation in active matter: First, there are oscillations due to translational inertia as in the passive case. These are still present in active matter, although their properties are modified by the coupling to polarization degrees of freedom. Second, there are oscillations that arise from the coupling of the density to the polarization. This mechanism is not present in fluids consisting of passive spherical particles. As can be seen in Figs.\ \ref{fig4}\fn{a}{ii} and \ref{fig4}\fn{c}{ii}, the damping of the green mode decreases for larger frequencies if we set $\epsilon=-1.5$, whereas the \ZT{passive} sound modes are more strongly damped in this case. Consequently, sound can propagate further into the medium via the second, \ZT{active} mechanism at higher frequencies at these parameter values.

We summarize our observations for sound propagation in active matter:
\begin{itemize}\itemsep3mm
\item For $v_0 = 0$, there is one mechanism for sound propagation (coupling of density and momentum), as is well known from passive fluids.
\item For active systems, stronger damping is required to ensure that the system is still stable against perturbations.
\item A second type of sound waves exists in the active case, arising from the coupling of density and polarization. (For $v_0 =0$, these waves only affect the polarization, not the density.)
\item For certain parameter values, waves of the first type propagate further into the medium at small frequencies, whereas waves of the second type propagate further at higher frequencies.
\end{itemize}

The last effect is of particular interest for possible applications, since it allows for the development of acoustic frequency filters: If one excites oscillations in the red mode, smaller frequencies will propagate further into the medium, i.e., the system constitutes a low-pass filter. On the other hand, if the green mode is excited, higher frequencies will experience less damping, such that we have a high-pass filter. The properties of the filters can be tuned by changing the system parameters as shown in \cref{fig4}.

\section{\label{conc}Conclusions}
We have shown how the presence of two relaxational time scales changes a general first-order soft matter model into a third-order spatiotemporal jerky dynamics. This general structure has then been used to derive a general active PFC model in which, in contrast to the standard case, relaxational and orientational time scales are both finite and different.

The resulting model as well as the method by which it was obtained allow to describe active matter systems in which inertia is important with greater accuracy than previous theories. Since our model is of third order in temporal derivatives (unlike usual theories which are of first or second order), it can be expected to show a number of interesting effects that are not present in other models. This is also plausible by comparison to the case of ordinary dynamical systems, where jerky dynamics allows for chaotic behavior that is not possible in second-order models.

From the dispersion relation, we were able to investigate the linear stability and the resulting phase diagram, showing that in active systems, where the phase boundaries are not determined by the free energy, changes of the damping coefficients can affect the linear stability of the fluid phase. The fluid state is destabilized by increased activity. Moreover, it is found that active fluids can exhibit two different mechanisms for sound propagation: waves arising from density-momentum coupling as in the passive case, which can propagate with small damping at low frequencies, and waves arising from density-polarization coupling, which can propagate further at high frequencies. This effect has potential applications in the development of acoustic frequency filters based on active fluids.

Possible extensions of this work include the consideration of translational-rotational coupling (as discussed for particles with general shape in Ref.\ \cite{WittkowskiL2011}). Moreover, one could consider spherical domains (as done previously for simpler PFC models \cite{PraetoriusVWL2018,MohammadiD2020}) or more general memory kernels.  

\begin{acknowledgments}
We thank Tobias Frohoff-H\"ulsmann, Svetlana V.\ Gurevich, Stefan J.\ Linz, Lukas Ophaus, Alice Rolf, and Uwe Thiele for helpful discussions. R.W.\ is funded by the Deutsche Forschungsgemeinschaft (DFG, German Research Foundation) -- WI 4170/3-1.
\end{acknowledgments}

\nocite{apsrev41Control}
\bibliographystyle{apsrev4-1}
\bibliography{control,refs}

\end{document}